\documentstyle[sprocl,epsfig]{article}
 
\begin{document}
\begin{titlepage}
\def\thefootnote{\fnsymbol{footnote}} 
 
\begin{center}
\mbox{ }
 
\vspace*{-3cm}
 
 
\end{center}
\begin{flushright}
\Large
\mbox{\hspace{10.2cm} hep-ph/0102342} \\
\mbox{\hspace{10.2cm} February 2001}
\end{flushright}
\begin{center}
\vskip 2.0cm
{\Huge\bf
Comparison of Higgs Boson Mass  \\
\vskip 0.2cm
and Width Determination of the \\
\vskip 0.6cm
LHC and a Linear Collider}
\vskip 2.5cm
{\LARGE\bf V.~Drollinger and A.~Sopczak}\\
\smallskip
\vskip 1cm
\Large University of Karlsruhe

\vskip 2.5cm
\centerline{\Large \bf Abstract}
\end{center}

\vskip 2.5cm
\hspace*{-1cm}
\begin{picture}(0.001,0.001)(0,0)
\put(,0){
\begin{minipage}{16cm}
\Large
\renewcommand{\baselinestretch} {1.2}
Two important properties of a Higgs boson are its mass and width. 
They may distinguish the Standard Model (SM) Higgs boson from Higgs bosons 
of extended models. 
We show results from a direct mass and width reconstruction for a 
Higgs boson mass range from 120 to 340 GeV/c$^2$.
The mass and width have been reconstructed from the 
$\rm H\rightarrow ZZ^\star\rightarrow \mu^+\mu^-\mu^+\mu^-$ 
reaction in an LHC simulation of the CMS detector.
The determined mass accuracy has been compared with that obtained from 
studies for a linear collider (LC). The mass precision from the 
latter studies is derived by scaling previous LC
simulation results according to the expected event rates.
For the Higgs boson width we compare a direct determination with
indirect methods and find good complementarity.

\renewcommand{\baselinestretch} {1.}
 
\normalsize
\vspace{1.0cm}
\begin{center}
{\sl \large
Talk at the Worldwide Workshop on Future $\rm e^+e^-$ Collider,
Chicago, November 2000, \\ to be published in the proceedings.
\vspace{-6cm}
}
\end{center}
\end{minipage}
}
\end{picture}
\vfill
 
\end{titlepage}
 
 
\newpage
\thispagestyle{empty}
\mbox{ }
\newpage
\setcounter{page}{1}

\title{Comparison of Higgs Boson Mass and Width Determination \\
       of the LHC and a Linear Collider}

\author{Volker Drollinger, Andr\'e Sopczak\footnote{speaker}}
\address{\it University of Karlsruhe}

\maketitle
\abstracts{
Two important properties of a Higgs boson are its mass and width. 
They may distinguish the Standard Model (SM) Higgs boson from Higgs bosons 
of extended models. 
We show results from a direct mass and width reconstruction for a 
Higgs boson mass range from 120 to 340 GeV/c$^2$.
The mass and width have been reconstructed from the 
$\rm H\rightarrow ZZ^\star\rightarrow \mu^+\mu^-\mu^+\mu^-$ 
reaction in an LHC simulation of the CMS detector.
The determined mass accuracy has been compared with that obtained from 
studies for a linear collider (LC). The mass precision from the 
latter studies is derived by scaling previous LC
simulation results according to the expected event rates.
For the Higgs boson width we compare a direct determination with
indirect methods and find good complementarity.}

\vspace*{-0.8cm}
\section*{Introduction}

The search for Higgs bosons is one of the main goals of
future colliders. After a discovery it will be very important
to establish the properties of the Higgs boson.
First, results from an LHC study are given for the gluon fusion
process where the Higgs boson mass is reconstructed by calculating 
the invariant mass of the four-muon final state.
The simulation is based on the CMS detector response 
parametrization \mbox{CMSJET}~\cite{cmsjet}.
The selection is outlined in Refs.~\cite{epj,CMS4MU} and it
suppresses effectively the background.

\begin{figure}[h!]
\vspace*{0.8cm}
\begin{minipage}{0.49\textwidth}
\vspace*{-0.5cm}
\centerline{\epsfig{file=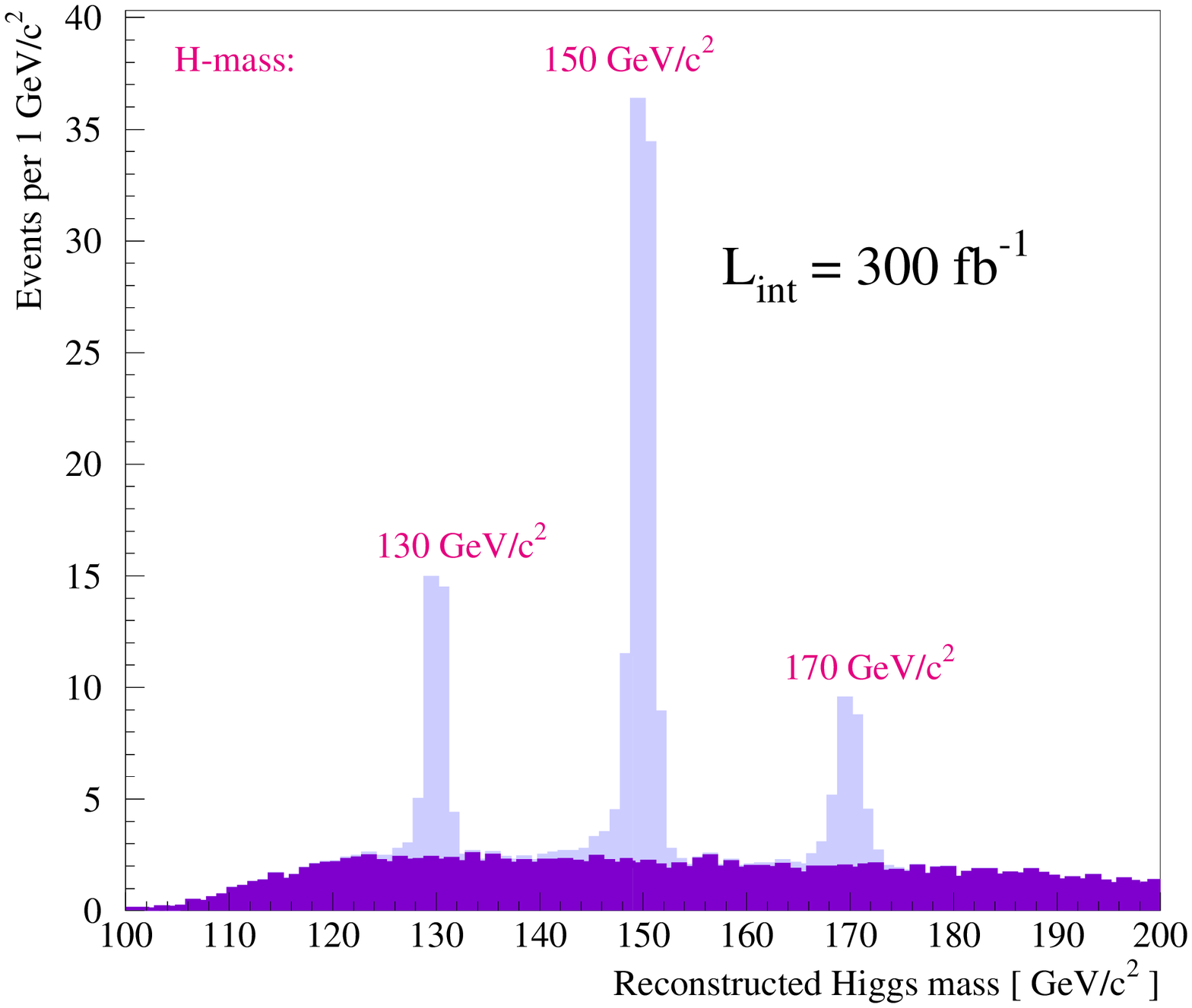,width=\textwidth}}
\vspace*{-0.3cm}
\caption{Expected signal and background in the CMS experiment
         for Higgs boson masses of 130, 150 and 170~GeV/c$^2$ 
         in the $\mu^+\mu^-\mu^+\mu^-$ channel for an integrated 
         luminosity of 300 $\rm fb^{-1}$.
         }
         \label{h4mu}
\end{minipage}
\hfill
\begin{minipage}{0.49\textwidth}
\vspace*{-0.8cm}
\centerline{\epsfig{file=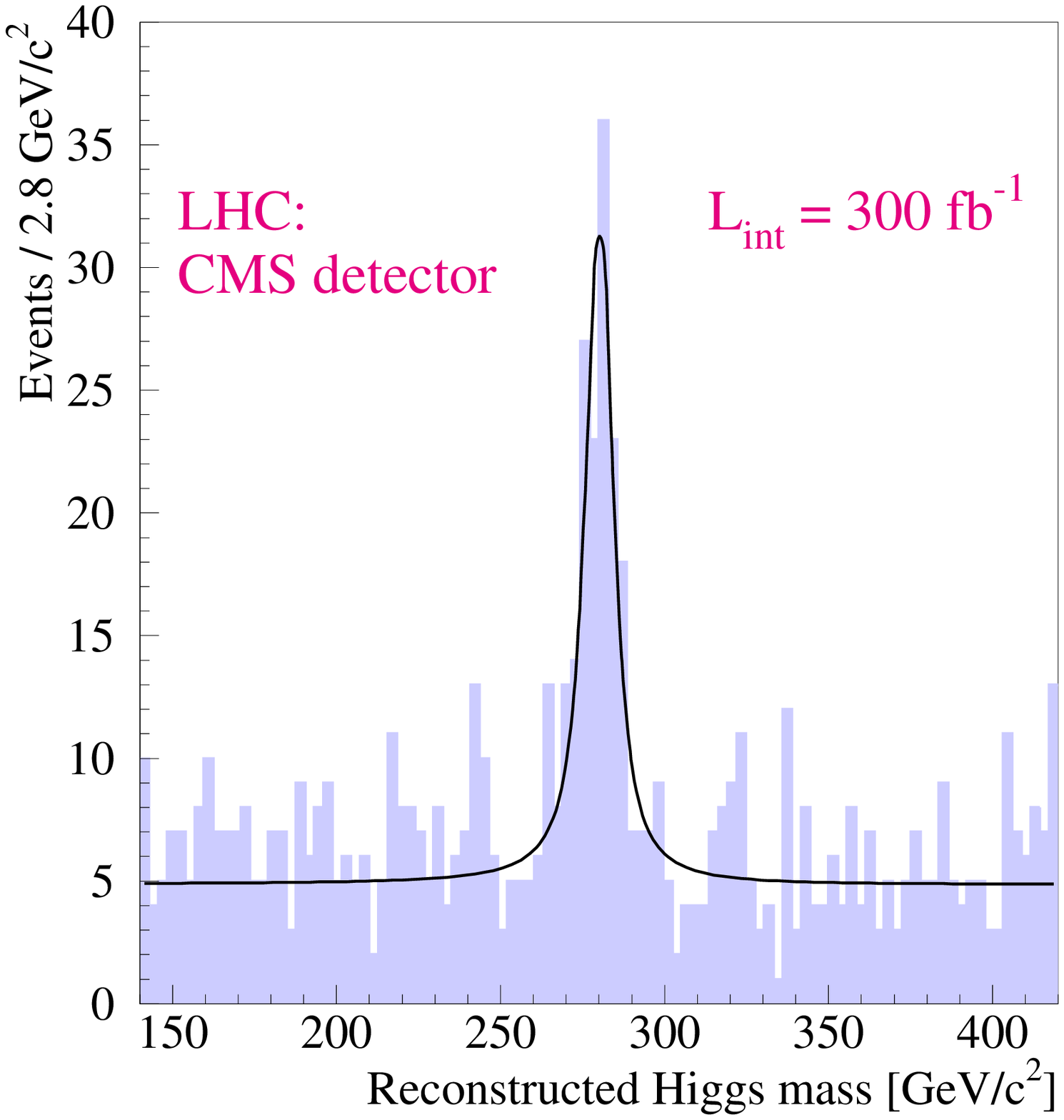,width=\textwidth}}
\vspace*{-0.5cm}
\caption
{Reconstructed Higgs boson mass in the 
 $\mu^+\mu^-\mu^+\mu^-$ channel as obtained from a CMS simulation
 (out of 100).
 For a simulated mass of 280 GeV/c$^2$, 
 $m_{\rm H}^{\rm rec} =$ 280.2 $\pm$ 0.6 GeV/c$^2$ 
 is reconstructed.
 \label{m160}
 }
\end{minipage}
\end{figure}

The invariant four-muon mass distribution is shown in 
Fig.~\ref{h4mu} for three different Higgs boson masses.  
The resulting peaks are clearly visible over the expected background which 
consists of ZZ, $\rm t\bar{t}$ and $\rm Zb\bar{b}$ events.
The number of background events is fairly constant and only about two
events per 1 GeV/c$^2$ for an integrated luminosity of 300 $\rm fb^{-1}$. 
The signal and background simulation is repeated 100 times
for each investigated mass, where the background distribution
is generated randomly according to the expected rate per mass bin.
The results of the 100 fits are averaged in order to obtain the most
likely outcome of the future experimental analysis.
Each invariant-mass distribution is fitted with a Gaussian below a Higgs boson
mass of 200~$\rm GeV/c^2$ and a Breit-Wigner distribution otherwise.
An example is shown in Fig.~\ref{m160} and all results are given in 
Table~\ref{tab:mass}. 
A description of the Higgs mass determination of the other LHC experiment,
ATLAS, can be found in Ref.~\cite{atlas}. 

{\small
\begin{table}[ht]
\begin{tabular}{c|cccccccccccc}
Mass     &  120&  140&  160&  180&  200&  220&  240&  260&  280&  300&  320&  340\\ \hline
$m_{\rm G}$
         &120.0&139.9&160.0&179.9&---&---&---&---&---&---&---&---\\
$\Delta m_{\rm G}$
         &0.38&0.13 &0.23 &0.24 &---&---&---&---&---&---&---&--- \\
$m_{\rm BW}$
         &---&---&---&---&199.9&219.9&239.8&259.9&279.9&299.8&320.1&340.3\\
$\Delta m_{\rm BW}$
         &---&---&---&---&0.14 &0.20 &0.28 &0.44 &0.58 &0.81 &1.1  & 1.5
\end{tabular}
\caption{ \label{tab:mass} Simulated and reconstructed masses for a Gaussian (G)
         and Breit-Wigner (BW) fit and their statistical errors for the 
         LHC study in the $\mu^+\mu^-\mu^+\mu^-$ channel.
         All values are given in GeV/c$^2$. 
         The masses and their statistical errors are the averages 
         from 100 simulated test experiments.}
\end{table}
\vspace*{-0.3cm}
}

\section*{Mass Reconstruction}

\begin{figure}[h!]
\begin{minipage}{0.49\textwidth}
\centerline{\epsfig{file=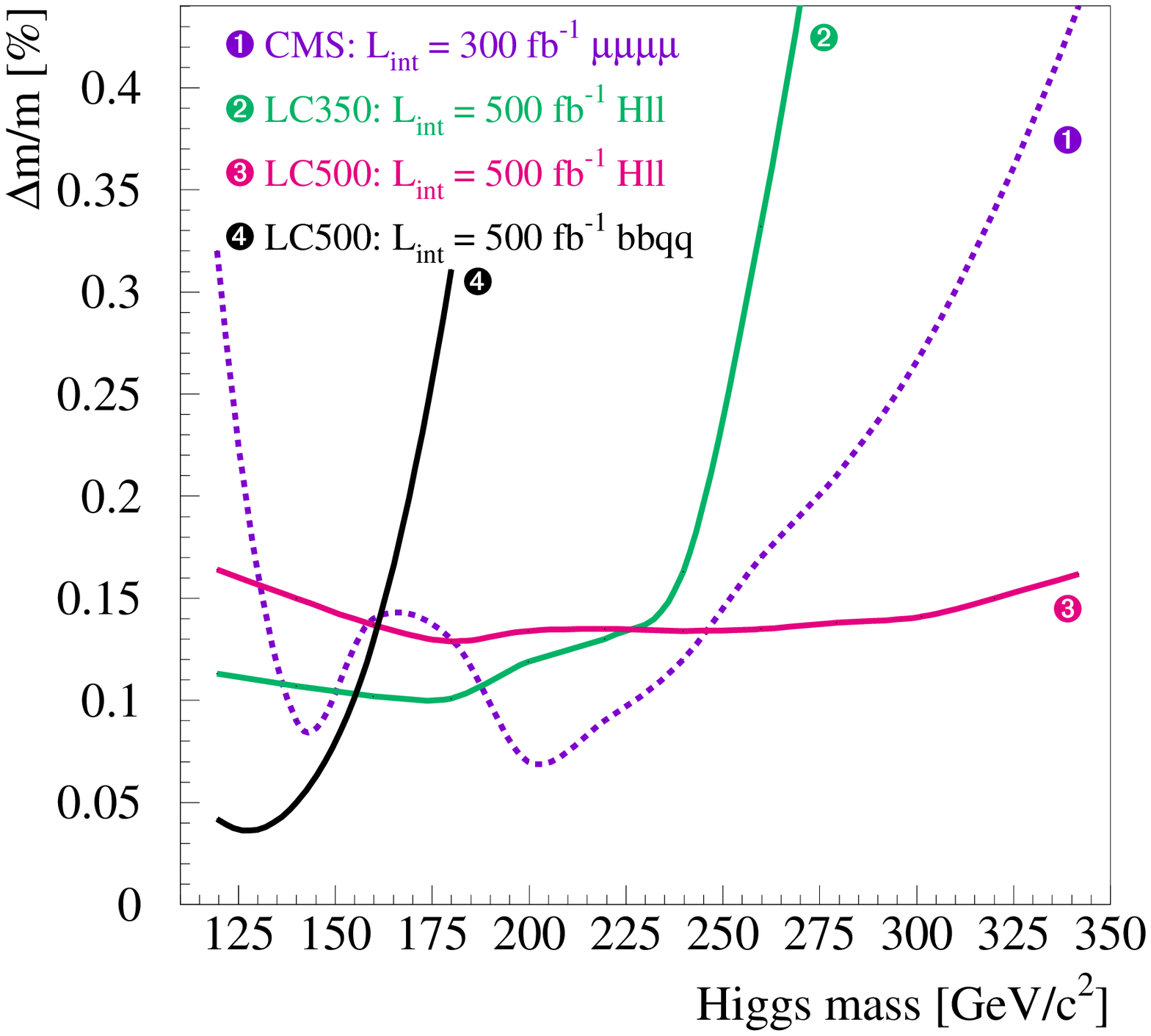,width=\textwidth}}
\vspace*{-0.7cm}
\caption{Relative error on the measurement of the SM Higgs mass
 versus mass for the LHC (CMS simulation) and an LC.
 \label{reso}
 }
\end{minipage}
\hfill
\begin{minipage}{0.49\textwidth}
\centerline{\epsfig{file=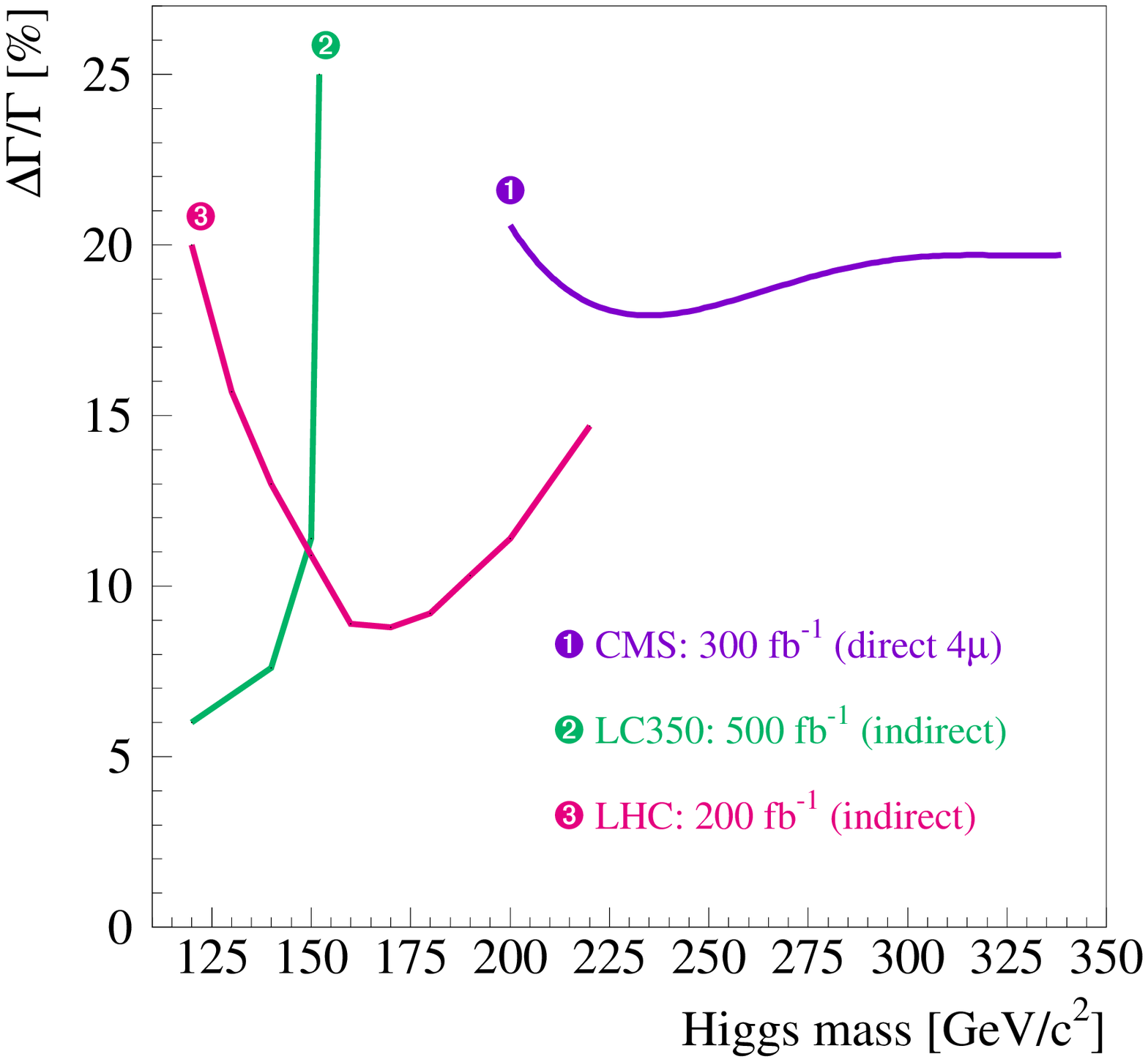,width=\textwidth}}
\vspace*{-0.7cm}
\caption{
 Relative error on the measurement of the SM Higgs width
 versus mass for the LHC (CMS simulation and indirect determination) 
 and an LC.
 \label{width}
 }
\end{minipage}
\end{figure}

This LHC study has been compared with two LC studies which have
been extrapolated to the mass range investigated here.
The first study has exploited the Higgsstrahlung production
$\rm e^+ e^- \rightarrow ZH$,
where the Z decays either to $\rm e^+e^-$ or $\mu^+\mu^-$. 
The reconstruction of the recoiling mass to the $\rm \ell^+\ell^-$ pair 
from the Z decay is used for the Higgs boson mass reconstruction. 
The two leptons from the Z boson decay can be reconstructed 
with a high precision~\cite{GarLoh}.
In a second study the ZH $\rightarrow$ $\rm b\bar{b}q\bar{q}$
channel was investigated~\cite{AJuste}. In this channel the Higgs 
boson mass was calculated from the invariant mass of the 
$\rm b\bar{b}$ system.
Beamstrahlung is included in all LC simulations and has 
only a small effect on the reconstructed mass distributions.
Details of the extrapolations are given in Table~\ref{tab2}.
Figure~\ref{reso} compares these LC studies with the Higgs 
boson mass determination for the LHC.

\begin{table}[ht!]
\begin{center}
\begin{tabular}{c|l}
  LHC: $\mu\mu\mu\mu$ & S: 120 - 340 GeV/c$^2$
(CMS detector response)\\ \hline
LC: H$\ell\ell$               & S: 120 - 160 GeV/c$^2$
$\sqrt{s} = $ 350 GeV \cite{GarLoh}\\
                      & E: 
$\Delta m_{H} = \sqrt{\sigma_0 / \sigma_{\rm H}} \times 
\Delta m_0$ for $\sqrt{s} = $ 350 and 500 GeV\\ \hline
  LC: $\rm b\bar{b}q\bar{q}$            & S: 
120 GeV/c$^2$ at $\sqrt{s} = $ 500 GeV \cite{AJuste}\\
                      & E: 
$\Delta m_{\rm H} = \sqrt{\rm BR_0(H \rightarrow b\bar{b}) 
/ BR_{H}(H \rightarrow b\bar{b})} \times \Delta m_0$
\end{tabular}

\caption{\label{tab2} Summary of simulations S and 
         extrapolations E. The variables with index zero are the 
values at the simulated Higgs boson mass of 120 GeV/c$^2$, and 
those with index H the values at higher Higgs boson masses
used for the extrapolation.}
\end{center}
\end{table}

\section*{Width Determination}

In addition to the Higgs boson mass determination,
the determination of the Higgs boson width may allow 
a Higgs boson of an extended model to be distinguished  
from a SM Higgs boson.
For the direct width determination at the LHC in the four-muon channel,
the detector response simulation and event selection is applied as
described before.

For masses below 200~GeV/c$^2$ the detector resolution dominates and a
Gaussian fit function is used, while for higher masses a Breit-Wigner 
distribution convoluted with the detector resolution is 
fitted and results are given in Table~\ref{tab:width}.
Figure~\ref{width} shows the expected accuracy, defined as the
error on the fit normalized to the natural width
of the SM Higgs boson (curve 1, CMS detector).
For a SM Higgs boson mass below about 180 GeV/c$^2$, the 
predicted width~\cite{HDECAY} becomes much smaller than the 
detector resolution.
In this range a direct width reconstruction is not possible
and only an upper limit can be derived.

{\small
\begin{table}[ht]
\begin{tabular}{c|cccccccccccc}
Mass      & 120& 140& 160& 180& 200& 220& 240& 260& 280& 300& 320& 340\\ \hline
$\Gamma_{\rm H}^{\rm SM}$
          &0.004 &0.008&0.077&0.63&1.4&2.3&3.4&4.8&6.5&8.5&10.9&13.8 \\
Det. resolution
          &0.35&0.63&0.63&0.76 &0.96 &1.1 &1.4 &1.6 &1.8 &2.0 &2.3 &2.4\\
$\sigma$&0.31&0.65&0.62&0.97&---&---&---&---&---&---&---&---\\
$\Delta\sigma$
        &0.27&0.12&0.22&0.24&---&---&---&---&---&---&---&---\\
$\Gamma$&--- &--- &--- &--- &1.6&2.7&4.1&4.8&7.5&9.5&12.8 &15.5 \\
$\Delta\Gamma$
        &--- &--- &--- &--- &0.29&0.41&0.64&0.85&1.3&1.6&2.2 &2.7\\
\end{tabular}
\caption{ \label{tab:width} 
         Simulated and reconstructed widths 
         and their statistical errors, and theoretical widths. 
         The simulated values are the averages for 100 
         test experiments for the LHC study in the 
         $\mu^+\mu^-\mu^+\mu^-$ channel. 
         $\sigma$ is the Gaussian fit result, 
         while $\Gamma$ is the fit result from a Breit-Wigner
         convoluted with the detector resolution.
         All values are given in GeV/c$^2$.}
\end{table}
}

In addition to the direct determination, results from
indirect determinations for an LC (curve~2) 
and for the LHC (curve~3) are shown in Fig.~\ref{width}.
The indirect methods are based on the W-fusion production
$\rm W^+W^-\rightarrow H$.
In the LC case~\cite{klaus}, and in contrast to the indirect LHC study,
only the decay $\rm H\rightarrow b\bar{b}$ is considered (curve~2).
The analysis of Ref.~\cite{dieter} (curve 3) makes 
use of several Higgs boson production processes.
In the mass region below 150 GeV/c$^2$, 
where the $\rm H\rightarrow W^+W^-$ branching fraction falls off rapidly,
the LC has a much higher sensitivity than the LHC.
The current direct and indirect studies for the LHC and an LC are complementary.

In future studies, the relative errors should improve further when,
for the LHC and LC, additional channels which use other
decay modes are investigated.

\section*{Conclusions}

We have investigated the accuracy with which the Higgs boson mass and width
could be measured at the LHC in the four-muon channel. These are compared 
with the results from LC studies extrapolated into the range 
120 - 340 GeV/c$^2$ in the case of the mass determination.
The accuracy of the mass measurement is limited by $\rm \sigma \times BR$ of 
the corresponding Higgs boson decay into ZZ and $\rm b\bar{b}$ for the 
$\rm H\rightarrow ZZ^\star \rightarrow \mu^+\mu^-\mu^+\mu^-$ (LHC) and
the $\rm e^+e^- \rightarrow HZ \rightarrow b\bar{b}q\bar{q}$ (LC) channels,
 respectively.
In contrast, determinations through the LC H$\ell\ell$ channel are independent
of the Higgs boson decay mode.
In the latter case the event rate depends only on the cross
section, which decreases slowly with the Higgs boson
mass and falls off rapidly at the kinematic production limit.
For\,a\,large mass range the mass precisions of the LHC and an
LC are comparable in\,the considered channels, and amount to\,about\,0.1\%.

For the Higgs boson width determination we show that direct
measurements and indirect methods cover the mass range 
120 - 340 GeV/c$^2$ with a relative error of about 10\%.
The overall precision will be higher when all Higgs decay modes
are investigated.


\clearpage
\section*{References}
\begin{thebibliography}{99}
\bibitem{cmsjet} 
S.~Abdullin, A.~Khanov, N.~Stepanov,
CMSJET version of September 15, 1998,
CMS TN/94-180.

\bibitem{epj} V.~Drollinger, A.~Sopczak, LC-PHSM-2000-037, IEKP-KA/2000-15,
              EPJdirect CN 1 (2000) 1.

\bibitem{CMS4MU} I.~Iashvili, R.~Kinnunen, A.~Nikitenko, D.~Denegri,
Study of the 
\mbox{$\rm H\rightarrow ZZ^\star\rightarrow 4\ell^\pm$} 
Channel in CMS, CMS TN/95-059;\\
V. Drollinger,
Studies of Higgs Boson Searches for 
the CMS Experiment, IEKP-KA/97-11, Diploma Thesis.

\bibitem{atlas} ATLAS Collaboration, 
Detector and Physics performance Technical Design Report, 
CERN/LHCC/99-15, vol. 2, Chapter 19.

\bibitem{GarLoh} P. Garcia-Abia and W. Lohmann, 
Measurement of the Higgs Cross Section and Mass 
with Linear Colliders, hep-ex/9908065.

\bibitem{AJuste} A.~Juste,
$M_{\rm H}$ Determination from Direct Reconstruction 
of ZH \mbox{($\rm Z \rightarrow q\bar{q}$)}, 
hep-ph/9912041.

\bibitem{HDECAY} A. Djouadi, J. Kalinowski and M. Spira,
HDECAY: a Program for Higgs Boson Decays in the 
Standard Model and its Supersymmetric Extension,
hep-ph/9704448.

\bibitem{klaus} K.~Desch, N.~Meyer, presentations at the LC Workshop,
                Obernai, France, Oct. 1999 and the LC Workshop,
                Padova, Italy, May 2000.

\bibitem{dieter} D.~Zeppenfeld, et al., Measuring the Higgs boson
                 coupling at the LHC, Phys. Rev. D 62 (2000) 13009.

\end{thebibliography}
\end{document}